\documentclass[reprint,superscriptaddress,amsmath,amssymb,pra]{revtex4-1}
\usepackage{amsmath}
\usepackage{bm}             
\usepackage{graphicx}
\usepackage{color}

\newcommand{\ZIB}{Zuse Institute Berlin, Takustra{\ss}e 7, D-14195 Berlin, Germany}
\newcommand{\JCM}{JCMwave GmbH, Bolivarallee 22, D-14050 Berlin, Germany}
\newcommand{\TUB}{Institute of Solid State Physics, Technische Universit\"at Berlin, D-10623 Berlin, Germany}

\DeclareMathAlphabet{\mathbfit}{OML}{cmm}{b}{it}

\begin{document}

\title{Numerical Investigation of Light Emission from Quantum Dots Embedded into On-Chip, Low Index Contrast Optical Waveguides}

\author{Theresa Hoehne}
\affiliation{\ZIB}
\author{Peter Schnauber}
\affiliation{\TUB}
\author{Sven Rodt}
\affiliation{\TUB}
\author{Stephan Reitzenstein}
\affiliation{\TUB}
\author{Sven Burger}
\affiliation{\ZIB}
\affiliation{\JCM}

\begin{abstract}
Single-photon emitters integrated into quantum optical circuits will enable new, miniaturized quantum optical devices.
Here, we numerically investigate semiconductor quantum dots embedded to low refractive index contrast waveguides. 
We discuss a model to compute the coupling efficiency of the emitted light field to the
fundamental propagation mode of the waveguide, and we optimize the waveguide dimensional parameters for
maximum coupling efficiency.
Further, we show that for a laterally cropped waveguide the interplay of Purcell-enhancement and optimized
field profile can enhance the coupling efficiency by a factor of about two. 
\end{abstract}
\maketitle  
\onecolumngrid
 {\footnotesize
\noindent This  is the  accepted  version  of  the  following  article:
T.\,Hoehne, et al., Phys.~Status Solidi B {\bf 256}, 1800437 (2019), 
which  has  been  published  in  final  form  at
https://doi.org/10.1002/pssb.201800437\,.
This  article  may  be  used  for  non-commercial purposes  in  accordance  with  the
Wiley  Self-Archiving Policy.\\
}
\twocolumngrid
\begin{figure*}[htb]
	\centering
	\def\svgwidth{0.75\textwidth}
	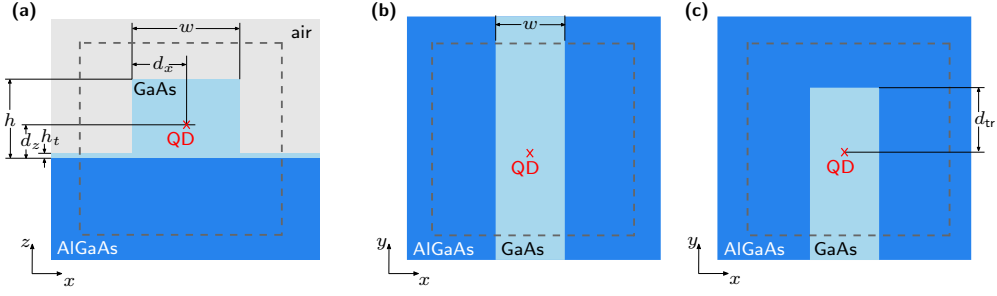
	\caption{
		Schematics of the investigated coupling geometries.
		a) Cross-section through the waveguide structure with geometry parameters
		(waveguide height, $h$, and width, $w$, top layer height, $h_\text{t}$, QD lateral and vertical placement, $d_x$, $d_z$).
		GaAs material (light blue), AlGaAs (dark blue), air (light gray) and embedded QD (red cross). 
		b) Top view of the model with a QD embedded into a waveguide which extends infinitely in $+y$ and $-y$ directions.
		c) Top view of the embedded QD with a waveguide which
                is truncated at distance $d_\text{tr}$ and extends infinitely in $-y$ direction.
		The dashed lines indicate the computational domain boundaries in the numerical simulations.
	}
	\label{fig:setup}
\end{figure*}

\section{Introduction}\label{introduction}

	Single-photon sources (SPSs) allow for numerous future technology applications that
	take advantage of well-defined quantum light states.~\cite{eisaman2011,shields07,senellart2017}
	SPSs enable fast information transfer via so-called \textit{flying} qubits, on a very compact and highly scalable level.
	Therefore, they represent a potential key element of quantum computers and quantum
	communication systems based on information transfer via single photons.~\cite{kiraz04,milburn09,bennett84}
	
	SPSs have been demonstrated using various technologies.~\cite{eisaman2011}
	Semiconductor quantum dot (QD) based realizations in principle allow for a 
	wide spectral range, stability and compatibility with on-chip technology.~\cite{aharonovich16,senellart2017}
	Integrating QDs into nano\-structured semiconductor environments with further
	functionalities allows for efficient devices with small footprint.~\cite{rengstl15,wang14,arcari14,kim17,dietrich16}
	Semiconductor ridge waveguides offer an opportunity to achieve good lateral guiding of
	emitted photons with low propagation loss.~\cite{hiroaki85}
	Embedding QDs directly into such waveguides allows for compact and efficient SPSs.~\cite{rengstl15,stepanov2015apl}
	
	In such structured local environments the Purcell effect plays an important role~\cite{purcell},
	i.e., the specific surrounding of an emitter modifies its spontaneous emission rate.~\cite{Reitzenstein2012,Canet-Ferrer2012}
	While the emission of photons into specific modes can be enhanced, the emission into
	other channels can be suppressed by means of tailoring the local mode density.~\cite{gregersen2012apl,stepanov2015apl}
	
	In this work we numerically investigate models of QDs embedded into on-chip wave\-guides.
	We concentrate on a material system which has recently been realized experimentally,~\cite{rengstl15,schnauber2018nl}
	where especially the deterministic integration of pre-selected QDs~\cite{schnauber2018nl}
	is highly advantageous for maximum coupling efficiency and for upscaling to larger and more complex circuits. 
	As the involved materials feature a low refractive index contrast, significant amounts of
    radiation couple to the continuum of substrate and free space modes.
    Therefore we naturally observe significantly smaller coupling efficiencies than what has
    been demonstrated using material systems with high refractive index contrasts.~\cite{stepanov2015apl,arcari14,laucht12}
    However, the material system under study is chosen since it can be reliably manufactured~\cite{rengstl15,schnauber2018nl}
    and can serve as a fault-tolerant solution for complex circuits where stable and
    feasible devices need to be fabricated with high repeatability.
		Firstly, the waveguide system presented here overcomes stability issues of suspended membrane structures. 
		Further, the good lattice constant match of waveguide core and cladding material of the proposed system 
		allows for the usage of established growing procedures.
		Complicated and error-prone transfer, printing or underetching techniques that
		are necessary for the fabrication of GaAs on insulator or free-standing platforms are avoided.
	
	We show that by engineering the Purcell factor and the overlap of the QD's emitted light field with guided modes,
	the coupling efficiency $\beta$	can be optimized.
	Specifically, lateral waveguide truncation is proposed for significantly enhanced and directed coupling.

\section{Setup and Methods}\label{section_setup}

	We investigate systems consisting of a single QD embedded into a ridge waveguide.
	A schematic of the device geometry model is shown in Figure~\ref{fig:setup}.
	The investigated material system is inspired by experimentally realized and demonstrated  structures.~\cite{rengstl15,schnauber2018nl} 
	The waveguide is formed by a gallium arsenide (GaAs) core of height $h$ and width $w$
        placed on a thick aluminium gallium arsenide	(Al$_{0.9}$Ga$_{0.1}$As) 
	substrate. 
	The small lattice mismatch between GaAs and 
			Al$_{0.9}$Ga$_{0.1}$As
		facilitates the fabrication of this material system.~\cite{manfra2014}
	A GaAs layer of thickness $h_t=50$\,nm prevents oxidation of the 
			Al$_{0.9}$Ga$_{0.1}$As,
		and the superspace is filled with air.
	
	The QD is embedded in the waveguide core at a height above the substrate, $d_z$,
	and at a lateral displacement from the waveguide edge, $d_x$, i.e., it is laterally centered for $d_x=w/2$. 
	Our model considers an indium gallium arsenide (InGaAs) QD emitting at a free-space wavelength of $\lambda_0=930$\,nm.
			As material refractive indices at this wavelength we use $n_\text{GaAs}=3.48$, 
		$n_{\text{Al}_{0.9}\text{Ga}_{0.1}\text{As}}=2.94$, and $n_\text{air}=1$. 
		The corresponding room temperature values~\cite{Jenkins1990} are extrapolated to a cryogenic operation 
		temperature of 5\,K based on ref.~\cite{Marple1964}.

	\begin{figure*}[htb]
		\centering
		\def\svgwidth{0.62\textwidth}
		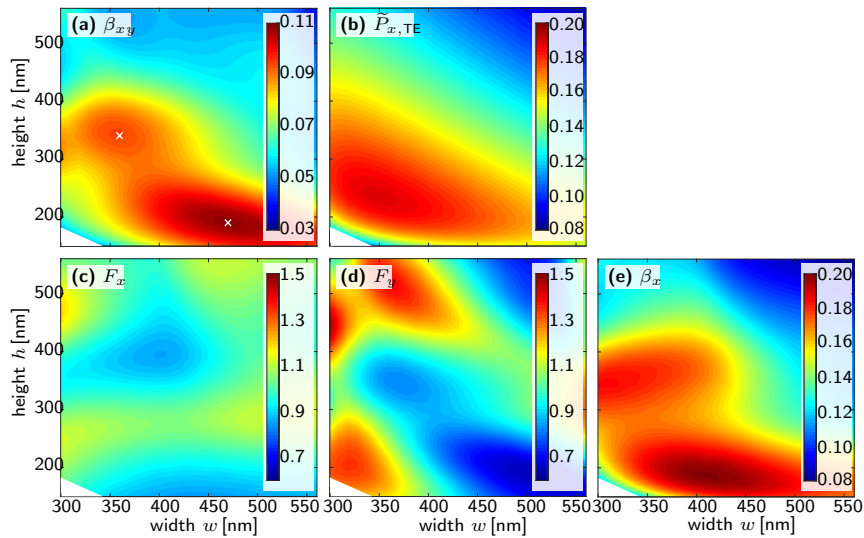
		\caption{%
			Impact of waveguide geometry parameters on the emission properties of the integrated QD.
			a) Coupling efficiency $\beta_{xy}$ as a function of waveguide width $w$ and height $h$.
	            b) Normalized mode coupling power
	            $\widetilde{P}_{x,\text{TE}}=P_{x,\text{TE}}/P_\text{bulk}$
		    and    c/d)
		    Purcell factors $F_{x/y}$ for $x/y$-polarized dipole emitters.
		        e)
	        Coupling efficiency of an $x$-polarized QD, $\beta_{x}$.
	            White crosses mark the waveguide parameter sets yielding maximum efficiency in two different regions in a),
	            specified as WG1 and WG2 in the following.
 		    }
		\label{fig:beta}
	\end{figure*}

	In a first set of simulations we consider a waveguide which extends infinitely to both directions, 
	$+y$, and $-y$, as shown in the top view in Figure~\ref{fig:setup}b.
	In a second set of simulations we investigate the impact of lateral waveguide truncation in $+y$-direction 
	on the coupling efficiency to modes propagating in the opposite direction.
	A top view of this setup is shown in Figure~\ref{fig:setup}c with the truncation distance labeled as $d_\textrm{tr}$.
	
	The QD emission is modeled as time-harmonic electromagnetic field radiated by oscillating electric 
	dipoles polarized in the $xy$-plane.~\cite{Schneider2018oe}
	The electric current density $\mathbfit{J}$ of the dipole source is given by 
	\begin{equation}\label{eq:source}
		\mathbfit{J}(\mathbfit{r},t) =
		\mathbfit{\cal{J}}(\mathbfit{r})\exp{(-i\omega t)} =
		\mathbfit{j}\delta(\mathbfit{r}-\mathbfit{r}_0)\exp{(-i\omega t)}
	\end{equation}
	with angular frequency $\omega=c_02\pi/\lambda_0$, speed of light in free-space $c_0$, 
	dipole position $\mathbfit{r}_0$ and amplitude vector $\mathbfit{j} =j(\cos\theta~\hat{x} + \sin\theta~\hat{y})$ 
	with unit vectors $\hat{x}$, $\hat{y}$ and constant amplitude $j$.
	In our model, the polarization angle $\theta$ in the $xy$-plane is randomly distributed: 
	$\theta\in[0,2\pi]$.
	Therefore the average electromagnetic field generated by the $xy$-polarized dipole source is described 
	by an incoherent superposition of fields emitted by an $x$-polarized $(\theta=0)$ and a $y$-polarized 
	$(\theta=\pi/2)$ dipole.~\cite{lethiec14}
	Correspondingly, the average total emission power of the source $P_{\text{tot}}$ can be written as 
	$P_{\text{tot}}= (P_x + P_y)/2$, where $P_{x(y)}$ is the power emitted by an $x(y)$-polarized dipole.
	
	The electric field $\mathbfit{\cal{E}}$ corresponding to the current density $\mathbfit{\cal{J}}$ 
	is found as a solution to the time-harmonic Maxwell's equation:
	\begin{equation}\label{eq:maxwell}
		\mathbfit{\nabla}\times\mu^{-1}\mathbf\nabla\times\mathbfit{\cal{E}}-\omega^2\varepsilon\mathbfit{\cal{E}} = i \omega\mathbfit{\cal{J}}
	\end{equation}
	with material permittivity $\varepsilon$ and permeability $\mu$.
	We assume $\mu=\mu_0$ and $\varepsilon = \varepsilon_r\varepsilon_0$ with vacuum permeability and permittivity, 
	$\mu_0$ and $\varepsilon_0$, and $\varepsilon_r=n^2$.
        
	In the following we denote the electric field solution to a scattering problem with source term $\mathbfit{\cal{J}}$ as $\mathbfit{\cal{E}}_s$,	while we denote	the electric field corresponding to time-harmonic eigenmodes defined on the waveguide cross-section at vacuum wavelength $\lambda_0$ as $\mathbfit{\cal{E}}_m$.~\cite{pomplun07}
	
	The impact on the spontanoues emission rate of the QD due to its specific surrounding is known as Purcell effect.~\cite{purcell}
	It is quantified by the dimensionless Purcell factor $F_s$,	that is the total emitted power $P_s$ of the dipole source in a structured area relative to its emission power $P_\text{bulk}$ in a homogenous environment with the same background material $n_b$.
	Here, the subscript $s$ denotes the source polarization direction, and the Purcell factor is
        calculated as follows:~\cite{Rao2008,Schulz2018,griffiths2014introduction}
		\begin{equation}\label{eq:purcell}
		F_s = \frac{P_s}{P_\text{bulk}} = 
		P_s~\left(\frac{n_b\mu_0\omega^2}{12\pi c_0}|\mathbfit{j}|^2\right)^{-1}\text{.}
	\end{equation}
        
			The scattered light field, $(\mathbfit{\cal{E}}_s,\mathbfit{\cal{H}}_s)$, 
		is partially coupled to the eigenmodes of the waveguide, $(\mathbfit{\cal{E}}_m,\mathbfit{\cal{H}}_m)$, 
		where $\mathbfit{\cal{H}}$ is the magnetic field strength which in the time-harmonic regime 
		can be expressed as the curl of the corresponding electric field: 
		$i\omega\mu\mathbfit{\cal{H}} = \nabla\times\mathbfit{\cal{E}}$.
		The power radiated into mode $m$ is computed by the overlap integral
		\begin{equation}\label{eq:p}
		P_{s,m} = \frac{\left|\frac{1}{2} \int_\Gamma\mathbfit{\cal{E}}_s \times \mathbfit{\cal{H}}_m \right |^2}
		{\left|\frac{1}{2} \int_\Gamma\mathbfit{\cal{E}}_m \times \mathbfit{\cal{H}}_m \right |^2}
		\Re{\left\{\frac{1}{2} \int_\Gamma\mathbfit{\cal{E}}_m \times \mathbfit{\cal{H}}^*_m\right\}}
		\text{,}
	\end{equation}
			where $\Gamma$ is the waveguide cross-section.

        The coupling efficiency $\beta_{s,m}$ for a source with polarization $s$ to a mode $m$ is then given 
	by the mode coupling power $P_{s,m}$ normalized to the emission power $P_s$ of the dipole:
	\begin{equation}\label{eq:beta}
		\beta_{s,m} = \frac{P_{s,m}}{P_{s}}\text{.} 
	\end{equation}
	
	Here, we investigate the coupling efficiency $\beta_{xy,\text{TE}}$ of photons emitted
	by an $xy$-polarized dipole to the fundamental TE mode of the GaAs waveguide structure in
	Figure~\ref{fig:setup}.
	The coupling of $xy$-polarized dipoles to TM modes is negligible.
	Following Equation~\eqref{eq:beta}, $\beta_{xy,\text{TE}}$ is determined as
	\begin{equation}\label{eq:betaX}
		\beta_{xy,\text{TE}} = \frac{P_{x,\text{TE}} + P_{y,\text{TE}}}{P_x + P_y}\text{.}%
	\end{equation}	
	
	In the following we simplify the notation by dropping the mode identifier,
	as here we are only interested in coupling to TE modes,
	i.e., $\beta_{xy} \equiv \beta_{xy,\text{TE}}$ and $\beta_{x} \equiv \beta_{x,\text{TE}}$.
	We especially aim for maximizing  $\beta_{xy}$ by tailoring the waveguide geometry and the
	dipole source placement within the waveguide.
	
	In order to numerically compute the electric field distributions $\mathbfit{\cal{E}}_s$ and 
	$\mathbfit{\cal{E}}_m$ and for evaluation of the integrals, Equation~\eqref{eq:p}, 
	we use the finite element method (FEM, solver JCMsuite).
	For obtaining accurate field distributions, higher-order finite-elements are applied,~\cite{pomplun07}
	as well as an adaptive mesh refinement and a subtraction-field formulation.~\cite{zschiedrich}
	The latter allows to overcome the poor regularity of $\mathbfit{\cal{E}}_s$ caused by the singularity at the dipole position.
	Transparent boundary conditions are implemented using error-controlled perfectly matched layers (PML).~\cite{Zschiedrich2006}
        The numerical parameters (polynomial order of the FEM ansatz functions, meshing) are chosen such 
	that an error of coupling efficiency values better than 
	0.1\,\%
	is reached.

        	\begin{figure*}[htb]
		\centering
		\def\svgwidth{0.99\textwidth} 
		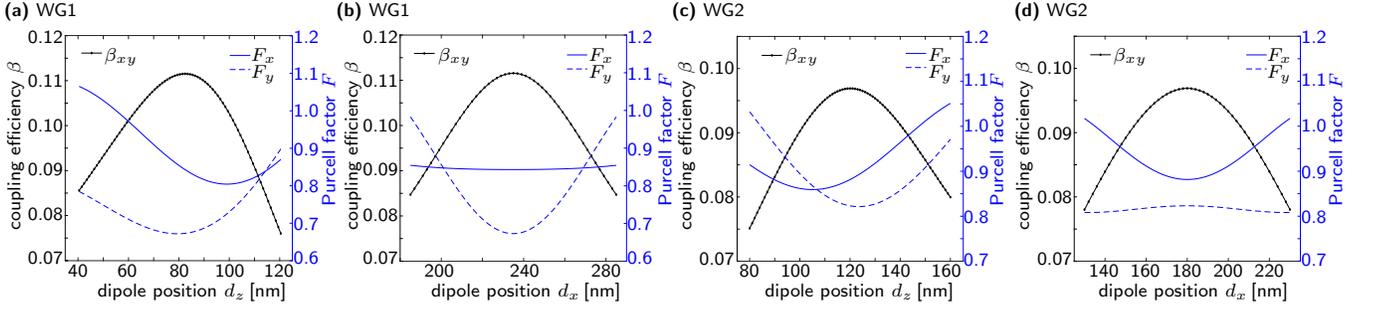
		\caption{Dependence of coupling efficiency $\beta_{xy}$ and Purcell factor $F_{x/y}$ on vertical dipole placement 
				$d_z$ (a/c) and horizontal dipole placement $d_x$ (b/d) for waveguide parameter set WG1 (a/b) 
				and waveguide parameter set WG2 (c/d).
		}
		\label{fig:dipole}
	\end{figure*}

\section{Results}

	First we investigate the coupling of light emitted from the embedded QD into the waveguide for a setup as depicted in Figure~\ref{fig:setup}a,b.
	With this, we search waveguide dimensions for which the coupling efficiency is maximized.
	For the given waveguide core height $h$ and width $w$ we compute the fundamental waveguide TE mode field distribution.
	The position of maximum electric field intensity of this mode is then used as the position where the QD is placed.
	Due to the symmetry of the problem the lateral placement is in the center of the waveguide, $d_x=w/2$.
	The vertical placement is roughly at $d_z\sim h/3$, depending on the specific set of  $h$ and $w$.
	It is lowered from the waveguide center due to the asymmetry in the superspace and substrate refractive indices.
     
	We scan the waveguide height and width over parameter ranges of $h\in\,$[150\,nm,\,560\,nm] and $w\in\,$[300\,nm,\,560\,nm], with a
	scan resolution of $\Delta h = \Delta w = 10\,$nm.
	For each parameter set we compute field distributions, Purcell factors and overlap integrals as described in Section~\ref{section_setup}.
	
	Calculated normalized mode coupling powers $\widetilde{P}_{x,\text{TE}}$,
	Purcell factors $F_x$, $F_y$ and coupling efficiencies $\beta_{xy}$ and $\beta_{x}$ as functions of $h$ and $w$ are shown in Figure~\ref{fig:beta}. 
	For mode coupling powers and coupling efficiencies, the values relate to one propagation direction ($-y$) of the waveguide.
	The white areas in the sub-plots of Figure~\ref{fig:beta} mark dimensions for which no guided eigenmodes are found.

	As a result, Figure~\ref{fig:beta}a shows the calculated coupling efficiency $\beta_{xy}$ for $xy$-polarized dipole sources.
		A maximum value of $\beta_{xy}=11\,\%$ per direction is reached for a waveguide with 
		geometry values $w_1=470$\,nm and $h_1=190$\,nm (white cross in Figure~\ref{fig:beta}a).
	In the following, the waveguide with these specific dimensions is referred to as WG1.
	The $\beta$ value of WG1 is comparable to calculations of ref.~\cite{rengstl15} where coupling of 
	$x$-polarized dipole emission into the fundamental TE and TM mode is considered.
    	Additionally, a second local maximum of $\beta_{xy}=9.5\,\%$ for parameters $w_2=360$\,nm and 
    	$h_2=340$\,nm is observed (second white cross in Figure~\ref{fig:beta}a).
    	We denote this specific waveguide as WG2 in the following.
	To better understand these results, we also investigate mode coupling powers and Purcell factors 
	calculated as in Equations~\eqref{eq:purcell} and \eqref{eq:p} for different source orientations.
	
	In Figure~\ref{fig:beta}b the mode coupling
        power $\widetilde{P}_{x,\text{TE}}$ for $x$-polarized dipoles  normalized to the dipole emission 
		power in a homogenous environment $P_\text{bulk}$ is plotted.
		$\widetilde{P}_{x,\text{TE}}$ increases with decreasing waveguide volume up to an optimum mode confinement. 
	In contrast $\widetilde{P}_{y,\text{TE}}$ 
		(not shown)
	is several orders of magnitude smaller 
	than $\widetilde{P}_{x,\text{TE}}$ due to the orthogonality of $y$-polarized dipoles and TE modes. 
	As expected, the coupling efficiency strongly depends on the dipole orientation. 
	
	The Purcell factors of $x$- and $y$-polarized dipole emitters inside the investigated device are depicted in 
		Figure~\ref{fig:beta}c,d. 
	Both $F_x$, $F_y$ are significantly impacted by changing the core dimensions of the waveguide. 
	Since the oscillation axis of $y$-polarized emitters is perpendicular to the direction of
	both dimension parameters, $F_y$ is more affected than $F_x$ by a variation of $h$ and $w$.
	The maximum value of coupling efficiency $\beta_{xy}$ is found at a position 
	where emission of the $y$-polarized source, $F_y$, is at a minimum.
	This is because its coupling efficiency $\widetilde{P}_{y,\text{TE}}$ is
	negligible. 
	The maximum value is also in a region where the Purcell effect for $x$-polarized sources is relatively small, 
	which can be understood by the fact that the variation of the values of $F_x$ in
        		Figure~\ref{fig:beta}c 
		is mainly due to enhancement and suppression of emission of radiation into channels 
	which do not couple to the waveguide mode. 
	The overall distribution of $\beta_{xy}$ depending on the waveguide cross-section size is therefore caused by 
	the complex interplay of varying mode volume and the Purcell effect.
	
   \begin{figure*}[htb]
		\centering
		\def\svgwidth{0.85\textwidth}
		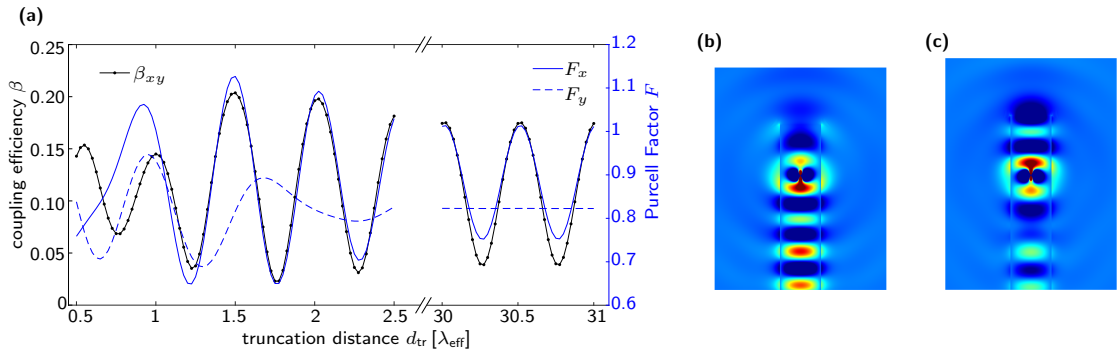
		\caption{
			Truncated waveguide (waveguide parameter set WG2) for enhanced directional coupling:
			a) Dependence of coupling efficiency $\beta_{xy}$  and Purcell factor
            $F_{x/y}$ on the distance between QD and waveguide termination, $d_\text{tr}$
             (in units of effective mode propagation wavelength, $\lambda_\text{eff}=298.7$\,nm).
 			b) Pseudo-color image of the induced $\Re(\mathbfit{\cal{E}}_x)$ field distribution for 
			$d_\text{tr}=1.5\times\lambda_\text{eff}$ in the $xy$-plane at the optimum $d_z$ position.
			A significant excitation of the fundamental mode propagating in the waveguide in $-y$ direction is visible.
			c) Same as (b) for  $d_\text{tr}=1.75\times\lambda_\text{eff}$ with minimal propagation mode excitation.
		}
		\label{fig:cut}
	\end{figure*}

		For a scattering problem with a purely $x$-polarized dipole source we observe another maximum for the coupling efficiency $\beta_{x}$, cf. Figure~\ref{fig:beta}e.
		Compared to the coupling of $xy$-polarized sources to the fundamental TE mode the global maximum value is increased to 
			$\beta_{x}=20\,\%$ and shifted to parameters $w=420$\,nm and $h=180$\,nm.
		This is a result of the strong dependence of $\beta$ on the dipole orientation, stated earlier.
	For our setup, a full suppression of $y$-polarized sources would thus lead to overall much higher coupling efficiencies. 	
	Besides, the $\beta$-values obtained for the GaAs/AlGaAs ridge waveguides are not comparable with 
	high index contrast realizations such as free-standing devices or others. 
	However, we are interested in the potential of devices for on-chip applications with high stability and reliablity when manufactured. 
	Here the proposed system outperforms the fragility of suspended realizations and overcomes 
	the need for additional transfers of the core to a different substrate.

	As next optimization step, we investigate the spatial dependence of the coupling
	efficiency by varying the vertical and horizontal dipole position $d_z$ and $d_x$ inside the waveguide, 
	cf. Figure~\ref{fig:setup}a.
	The results of this study show a parabolic trend of $\beta_{xy}$ with varying $d_z$
		(Figure~\ref{fig:dipole}a,c)
	and $d_x$ 
		(Figure~\ref{fig:dipole}b,d).
	For WG1, the optimum $z$-position is 
		12\,nm 
	above the TE mode maximum $z$-position. 
	However, only a slight increase of $\beta_{xy}$ by 
		0.3\,\% 
	is observed. 
	For WG2, the highest coupling efficiency corresponds to a vertical position of 
		$d_z=120$\,nm. 
	Here, the optimum $z$-position deviates by 
		13\,nm 
	from the TE mode maximum $z$-position. 
	The correction causes an increase of the coupling efficiency $\beta_{xy}$ by 
		0.3\,\%. 
	
	The Purcell factors $F_x$ and $F_y$ are roughly parabolic functions of $d_z$ but of opposite 
	sign than $\beta_{xy}$,	cf.~Figure~\ref{fig:dipole}a. 
		Their minima are at values of $d_z$ which differ from the TE mode maximum $z$-position,
		that is the position of highest mode coupling power $P_{x,\text{TE}}$ (not shown).
	So, the optimal vertical position for maximized $\beta_{xy}$ deviates from the position of 
	the TE mode maximum as a consequence of the Purcell effect.
	The horizontal dipole position $d_x$ of maximum coupling efficiency matches the
	waveguide center $d_x=w_2/2$ due to the device symmetry along the $z$-axis, cf. Figure~\ref{fig:dipole}b.
	As a result of the symmetric waveguide setup, the Purcell factors as functions of the 
	horizontal dipole position $d_x$ are also symmetric around the waveguide center.
	
	The results shown in Figure~\ref{fig:dipole} can also be used for quantifying the impact 
	of fabrication intolerances with regard to placement errors of the quantum dot within the waveguide.
	In combination with a deterministic integration of QDs \cite{gschrey2015,schnauber2018nl} 
	and high alignment accuracies, e.g. as of $24$\,nm realized in ref.~\cite{gschrey2015}, 
	an ideal coupling of the QD emission to waveguide modes can be achieved.
	In comparison, standard procedures, where high $\beta$ values can be reached,
    have low yield in successful processed devices.\\
     
    For an infinitely extended waveguide, on average, the same number of photons will be 
    coupled to modes, propagating in $-y$- and $+y$-direction each. 
	However, in typical applications, emission to a specific direction is required. 
	To increase the coupling efficiency to a specific direction, 
	we investigate the impact of the truncation of the waveguide at a distance $d_\text{tr}$, 
	which is the lateral distance between the dipole position and the waveguide-air interface. 
	A schematic view of this situation is illustrated in Figure~\ref{fig:setup}c.
	In a set of simulations we have varied the truncation distance $d_\text{tr}$ for
	waveguide setups WG1 and WG2, with optimized source positions. 
	We observe a strong impact on the coupling efficiency $\beta_{xy}$, with values between 
	    17.5\,\% (20.5\,\%) and 5\,\% (2\,\%) for WG1 (WG2).
	Figure~\ref{fig:cut}a shows the oscillating behaviour of $\beta_{xy}$ and $F_{x/y}$ as 
	functions of $d_\text{tr}$ for WG2. 
	    For large $d_\text{tr}$, these amplitudes converge to values of 15.5\,\% (17.5\,\%) 
	    and 7.5\,\% (4\,\%) for WG1 (WG2).
    
	The maximum 
		$\beta_{xy}=20.5\,\%$
	is reached for waveguide design WG2 with a cross-section size of 
		$360\,\text{nm}\times 340\,\text{nm}$
	and a truncation distance of $d_\text{tr}=1.5\times\lambda_\text{eff}$.
	The corresponding electric field distribution is visualized in Figure~\ref{fig:cut}b.
	Figure~\ref{fig:cut}c shows the field distribution (at equal color scale) for a 
	truncation distance such that the coupling efficiency is at its minimum.
	Clearly, the waveguide mode propagating outwards at the lower computational domain boundary is
	at significantly lower intensity than in Figure~\ref{fig:cut}b.
	For WG1 with cross-section $470\,\text{nm}\times190\,\text{nm}$, 
	$\beta_{xy}$ can be	tuned up to 
		$17.5\,\%$ with a truncation distance of  $d_\text{tr}=1.05\times\lambda_\text{eff}$.
	
	The observed oscillations can be understood from the following: 
	The emission in propagation direction is magnified when the standing wave field 
	amplitude at the dipole position is high,
	which means the source is at an antinode of the interference pattern of the 
	emitted field,~\cite{miller99} yielding a high Purcell factor. 
	Vice versa, the emission is attenuated when the point source is at a node position of
	its emitted field. 
	Consequently, local maxima can be found at even multiples of $\lambda_\text{eff}/4$,
	whereas minima are at odd multiples of $\lambda_\text{eff}/4$, with 
		$\lambda_\text{eff}=\lambda_0/\Re({n_\text{eff,TE}})=298.7$\,nm
	and $n_\text{eff,TE}$ is the effective refractive index of the 
	fundamental TE eigenmode of the waveguide WG2.
	For a polarization direction without significant emission into 
	the propagation direction, e.g. $y$-polarized dipoles, this effect is vanishing small.
	
	Hence, truncating the waveguide in one propagation direction can lead to directed emission 
	into the other with almost doubled coupling efficiency as a result of the tailored geometry. 
		
	\section{Conclusion}
	
	By tailoring the cross-section of an integrated GaAs waveguide device and its lateral geometry,
	a coupling efficiency of
		$\beta_{xy}=20.5\,\%$
	for collecting photons,
	emitted from an embedded, unpolarized SPS in a directed waveguide mode is obtained in numerical simulations.
	We first optimized the waveguide cross-section such that coupling into the fundamental TE mode is most efficient. 
	By varying the horizontal and vertical emitter position, we investigated the spatial dependence
	of $\beta_{xy}$ and the Purcell factor. 
	Finally, lateral truncation of the waveguide core enables almost a doubling of $\beta_{xy}$
	compared to an infinitly prolonged waveguide.
	This concept can naturally be applied to other material and waveguide systems for enhanced directional coupling. 
	Additionally, we find a strong dependence of $\beta_{xy}$ on the dipole orientation
	indicating that a full suppression of $y$-polarized sources leads to overall much higher
	coupling efficiencies exceeding
		35\,\%
	for the investigated truncated GaAs/AlGaAs waveguide design. 
	
	With this, we expect further enhancement possibilities of $\beta_{xy}$ through additional lateral structuring,
	and through usage of dedicated SPS with directed emission properties.~\cite{davanco17nc} 
	This investigation serves for the design of integrated waveguides~\cite{schnauber2018nl,davanco17nc,schwartz2016,joens2015} 
	and	also allows to estimate deviations from optimum coupling values due to fabrication tolerances. 
	This will help optimizing the performance of compact optical devices with integrated, 
	efficient SPS, which will pave the way for integrated quantum information processing.
        \\
        \\
        {\bf Acknowledgment} --
The authors gratefully acknowledge Lin Zschiedrich for helpful
discussions. This work has received financial support from the Deutsche
Forschungsgemeinschaft (DFG) within the Sonderforschungsbereich
SFB 787, project B4 (CRC~787-B4) and project C12 (CRC~787-C12). This
work is partially funded through the project 17FUN01 (BeCOMe) within
the Programme EMPIR. The EMPIR initiative is co-founded by the
European Union's Horizon 2020 research and innovation program and
the EMPIR Participating Countries.

\providecommand{\WileyBibTextsc}{}
\let\textsc\WileyBibTextsc

\end{document}